\documentclass[11pt,prd,reprint,twocolumn,notitlepage,nofootinbib]{revtex4-1}

\usepackage{amsmath}
\usepackage{graphicx}

\newcommand{\snr}{\mathrm{SNR}}
\newcommand{\gr}{\mathrm{GR}}
\newcommand{\ag}{\mathrm{AG}}
\newcommand{\ff}{\mathrm{FF}}

\newcommand{\thr}{\mathrm{thr}}

\newcommand{\Oc}{\mathcal{O}}
\newcommand{\odds}{\Oc}

\newcommand{\apjl}{ApJ Lett.}

\begin{document}

\title{Testing General Relativity with gravitational waves: a reality check}
\author{Michele Vallisneri}
\address{Jet Propulsion Laboratory, California Institute of Technology, Pasadena CA 91109}

\begin{abstract}
The observations of gravitational-wave signals from astrophysical sources such as binary inspirals will be used to test General Relativity for self consistency and against alternative theories of gravity. I describe a simple formula that can be used to characterize the prospects of such tests, by estimating the matched-filtering signal-to-noise ratio required to detect non-General-Relativistic corrections of a given magnitude. The formula is valid for sufficiently strong signals; it requires the computation of a single number, the \emph{fitting factor} between the General-Relativistic and corrected waveform families; and it can be applied to all tests that embed General Relativity in a larger theory, including tests of individual theories such as Brans--Dicke gravity, as well as the phenomenological schemes that introduce corrections and extra terms in the post-Newtonian phasing expressions of inspiral waveforms.
The formula suggests that the volume-limited gravitational-wave searches performed with second-generation ground-based detectors would detect alternative-gravity corrections to General-Relativistic waveforms no smaller than 1--10\% (corresponding to fitting factors of $0.9$ to $0.99$).
\end{abstract}

\maketitle

\section{Introduction and main results}
\label{sec:intro}

The possibility of performing high-precision tests of General Relativity (GR) in its dynamical, strong-gravity regime \cite{2006LRR.....9....3W} is perhaps the most exciting prospect of the budding field of gravitational-wave (GW) astronomy \cite{2009LRR....12....2S}. Several authors have carried out detailed analyses of such tests for both ground-based and space-based GW detectors \cite{1994PhRvD..50.6058W,1995PhRvD..52.2089K,1998PhRvD..57.2061W,2002PhRvD..65d2002S,2003CQGra..20S.219W,2004CQGra..21.4367W,2005CQGra..22S.943B,2005ApJ...618L.115J,2005PhRvD..71h4025B,2009CQGra..26o5002A,2009PhRvD..80d4002S,2010PhRvD..82l2001K,2010PhRvD..81f4008Y,2011PhRvD..84j1501B,2012PhRvD..85b4041M,2012PhRvD..85l2005B,2006CQGra..23L..37A,2006PhRvD..74b4006A,2010PhRvD..82f4010M,2011arXiv1108.1826H,2009PhRvD..80l2003Y,2011PhRvD..83h2002D,2011PhRvD..84f2003C}; by and large, the tests proposed so far belong in two classes.

In the first, GR is tested against specific alternative theories, such as scalar--tensor or massive-graviton theories, which recover GR for particular value of one or more additional parameters, such as the Brans--Dicke coupling constant, or the graviton mass \cite{1994PhRvD..50.6058W,1995PhRvD..52.2089K,1998PhRvD..57.2061W,2002PhRvD..65d2002S,2003CQGra..20S.219W,2004CQGra..21.4367W,2005CQGra..22S.943B,2005ApJ...618L.115J,2005PhRvD..71h4025B,2009CQGra..26o5002A,2009PhRvD..80d4002S,2010PhRvD..82l2001K,2010PhRvD..81f4008Y,2011PhRvD..84j1501B,2012PhRvD..85b4041M,2012PhRvD..85l2005B}. Thus, the strength of the tests is characterized by the accuracy with which the alternative-theory parameters can be measured and either found to be consistent with GR, or to deviate from it.

In the second class of tests, GR is tested for self-consistency by treating some of the coefficients in the post-Newtonian (PN) expansion of the phasing as free variables rather than deterministic functions of the source parameters, and verifying whether the recovered values are consistent with GR predictions \cite{2006CQGra..23L..37A,2006PhRvD..74b4006A,2010PhRvD..82f4010M,2011arXiv1108.1826H}. The strength of these tests is characterized by the amplitude of the deviations from GR that could be discerned in the PN coefficients.
More general tests are possible with the parametrized post-Einstein (ppE) formalism \cite{2009PhRvD..80l2003Y,2012PhRvD..86b2004C}, which, in addition to modifying the PN coefficients, adds extra terms to the PN amplitude and phasing and to the merger and ringdown waveforms, and recovers individual alternative theories for specific forms of the extra terms.

As advocated in \cite{2011PhRvD..83h2002D,2011PhRvD..84f2003C}, GR-by-GW tests find a more satisfying formulation in Bayesian model selection \cite{Jaynes,2005blda.book.....G}, which compares the \emph{Bayesian evidence}, given the observed data $s$, for the alternative-theory/modified-GR scenario (henceforth ``AG,'' for ``alternative gravity'') and for the Einstein-GR hypothesis. Model selection was applied to the PN consistency tests in Refs.\ \cite{2011PhRvD..83h2002D,2012PhRvD..85h2003L,2012JPhCS.363a2028L}, and to ppE inspiral waveforms in \cite{2011PhRvD..84f2003C}. (For a comprehensive discussion of model selection in the context of GW detection, rather than GR tests, see also Refs.\ \cite{2008CQGra..25r4010V,2008PhRvD..78b2001V,2008PhRvD..77h2002U,2009PhRvD..80f3007L}.) To wit, in model selection we compute the Bayesian \emph{odds ratio}
\begin{equation}
\Oc = \frac{P(\ag|s)}{P(\gr|s)} = \frac{
P(\ag) \int p(s|\theta^{i,a}) \, p(\theta^{i,a}) \, d\theta^{i,a}
}{
P(\gr) \int p(s|\theta^{i}) \, p(\theta^{i}) \, d\theta^{i}
},
\end{equation}
where $P(\ag)$ and $P(\gr)$ are the prior probabilities assigned to the AG and GR hypotheses; $\theta^i$ and $\theta^a$ are the source parameters (masses, spins, etc.) and additional AG parameters, respectively; $p(s|\theta)$ is the likelihood of the observed data $s$ given $\theta$; and $p(\theta)$ is the prior probability distribution for $\theta$.\footnote{In this paper we forgo annotating probabilities with the customary conditional dependence on ``all other'' assumptions, usually denoted as $I$.} The odds ratio describes the degree to which we should prefer one hypothesis over the other after having observed the data, and it incorporates the Bayesian law of parsimony (a.k.a.\ Occam's razor)---although models with additional parameters will always fit the data better, they will be relatively disfavored by the improbability that more parameters assume particular values in their prior ranges \cite{Jaynes,2005blda.book.....G}.

A cogent way of understanding the statistical significance of odds ratios is to set up a \emph{decision scheme} based on the value of $\odds$ \cite{2008CQGra..25r4010V,2012JPhCS.363a2028L}. Namely, we declare that we have detected AG whenever $\odds$ is greater than a set threshold $\odds_\thr$. We set $\odds_\thr$ by requiring a given \emph{false-alarm rate} $F$: this is the fraction of observations in which the underlying signal is GR, but $\odds$ happens to pass the threshold. $F$ gets smaller the more averse we are to falsely claiming AG detection, and its choice in practice should be guided by the prior $P(\ag)$. Now, for a given $\odds_\thr$, the \emph{efficiency} $E$ of detection is the fraction of observations in which the underlying signal is AG, and $\odds$ passes the threshold, so AG is detected correctly.\footnote{The performance of decision schemes is characterized by their \emph{receiver operating characteristic} $E(F)$ \cite{2008arXiv0804.1161S}. Note that the term ``fraction,'' used above in defining $F$ and $E$, is ideally the fraction of an infinite number of observations of the same GW signal immersed in different realizations of noise. This characterization of decision schemes is therefore a \emph{frequentist} statement (about the Bayesian statistic $\odds$), but one that this Bayesian author finds very reasonable.}
A way of understanding the strength of a test of GR is then to choose a reasonably low $F$ (say, $10^{-4}$) and ask how strong an AG effect and how loud a GW signal we would need to detect AG with reasonably high $E$ (say, $1/2$, but it turns out in practice that $E$ rises sharply after that).

In Ref.\ \cite{2011PhRvD..84f2003C}, Cornish and colleagues point out that the odds ratio for AG over GR grows with the signal-to-noise ratio (henceforth, SNR) of the \emph{residual} obtained after the best-fit GR waveform has been subtracted from the data; thus, alternative models that are not fit well by varying the GR parameters can be detected more easily than models that are. Indeed, Cornish and colleagues show that in the limit of large signal $\snr$ and small AG deviations the logarithm of the odds ratio scales as $(1 - \ff) \, \snr^2$, with $\ff$ the \emph{fitting factor} \cite{1995PhRvD..52..605A} between the GR and AG waveforms:
\begin{equation}
\label{eq:ff}
\ff(\theta_\ag) = \max_{\theta_\gr} \frac{\bigl( h_\gr(\theta_\gr), h_\ag(\theta_\ag) \bigr)
}{
|h_\gr(\theta_\gr)| \, |h_\ag(\theta_\ag)|
}.
\end{equation} 
Here $h_\gr(\theta_\gr)$ and $h_\ag(\theta_\ag)$ are the GR and AG waveform families (so $\theta_\gr \equiv \theta^i$ and $\theta_\ag \equiv \theta^{i,a}$), and $(\cdot,\cdot)$ is the standard noise-weighted inner product, such that the sampling probability of a Gaussian-noise realization $n$ is $\propto e^{-(n,n)/2}$, and the optimal matched-filtering SNR of an observed signal $h$ is its norm $|h| \equiv (h,h)^{1/2}$ (see, e.g., \cite{1994PhRvD..49.2658C}).
In the FF, the parameters $\theta_\ag$ are fixed by the AG waveform contained in the data, and the inner product is maximized over $\theta_\gr$. The FF is by definition independent of SNR, and it tends to one when the AG corrections vanish or can be completely reabsorbed by varying $\theta_\gr$.

In this paper I formalize and generalize this scaling statement by deriving the full expression of the odds ratio for the AG and GR hypotheses, in the limit of large SNR; the result is valid when AG embeds GR, which is the case for all classes of tests discussed above\footnote{I thank Curt Cutler for pointing out that this is true also for the PN-coefficient tests.} (see Sec.\ \ref{sec:oddsratio}). Moreover, I derive the decision-scheme statistics for the resulting $\odds$, and show that the efficiency $E(F)$ is a remarkably simple function [Eq.\ \eqref{eq:roc}, a combination of the error function and its inverse] of the \emph{effective} signal-to-noise ratio $\snr \sqrt{1 - \ff}$ (see Sec.\ \ref{sec:decision}). No other information about the waveforms is needed.

Thus, AG detection by model comparison allows us to characterize very generally both kinds of tests discussed above, by computing the \emph{SNR required to positively detect an AG correction as a function of its FF}. Given the sensitivity curve of the detector and the projected detection rates for a source class, we can then derive the magnitude of the AG corrections that we expect to be able to constrain in our observation campaigns. The FF can be computed from the GR Fisher matrix using the formulas of Ref.\ \cite{2007PhRvD..76j4018C}, or directly by maximizing the normalized product \eqref{eq:ff} over $\theta_\gr$.
\begin{figure}
\includegraphics[width=0.48\textwidth]{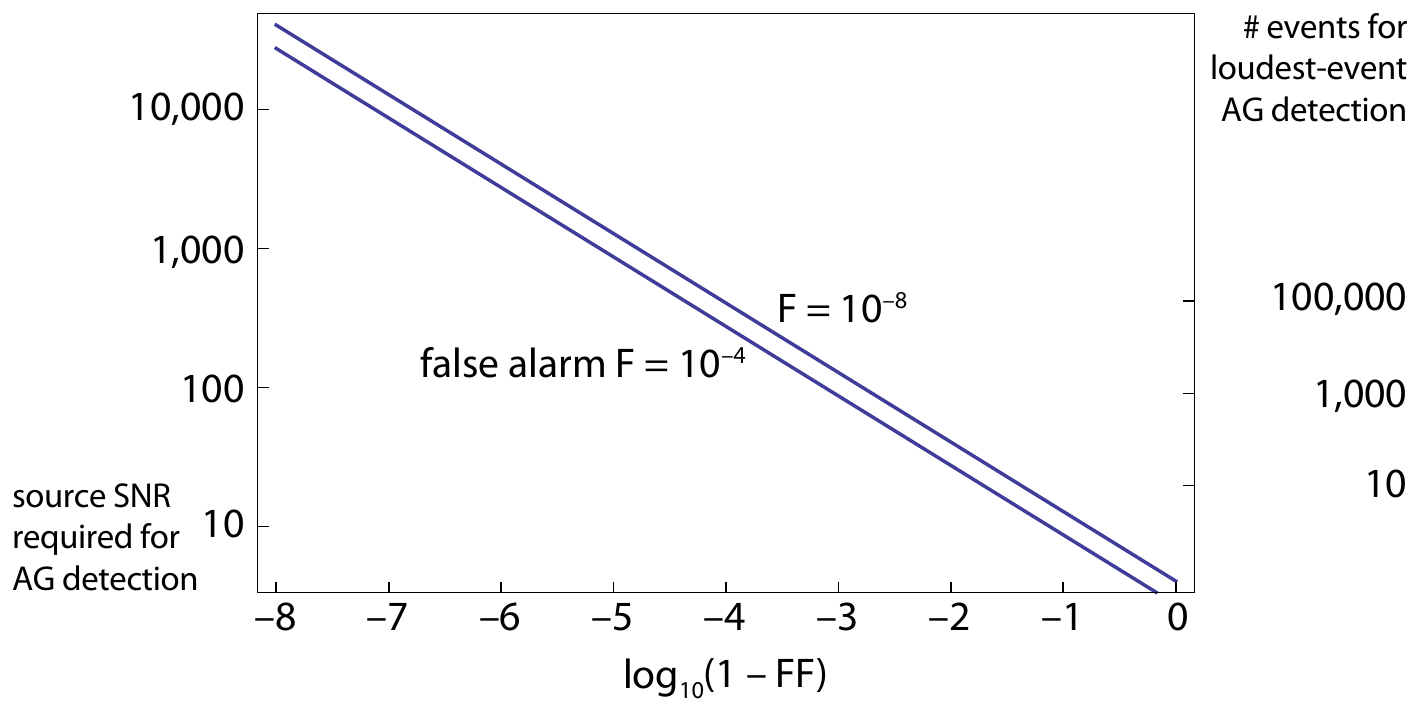}
\caption{\label{fig:reqsnr}SNR required for AG detection with efficiency $E = 1/2$, with false-alarm probability $F = 10^{-4}$ and $10^{-8}$, as a function of FF. The right-side vertical axis shows the number of events required in a volume-limited search with detection threshold of 8 to yield a loudest event with the (median) SNR on the left-side vertical axis.}
\end{figure}

The AG-detection SNR is shown in Fig.\ \ref{fig:reqsnr} for $F = 10^{-8}\mbox{--}10^{-4}$, and it is a rather exacting function of $1 - \ff$. For the typical observations produced by volume-limited searches, which have SNRs at the event-detection threshold ($\simeq 8$), only 10\% AG corrections ($1 - \ff = 0.1$) would be detectable, although for most waveform families the strong-signal approximation would not be appropriate at such low SNRs \cite{2008PhRvD..77d2001V}. The required $\snr$ grows roughly threefold for each decade of $1 - \ff$, to $\snr \gtrsim 30$ for 1\% effects, $\snr \gtrsim 100$ for one-in-a-thousand effects, and $\snr \gtrsim 1,000$ for one-in-ten-thousand effects. 

We can also compute easily the total volume-limited detection rates that would yield one event strong enough (on the median) to detect AG corrections with a given $1 - \ff$ (see Sec.\ \ref{sec:decision}); these are shown on the right-side vertical axis of Fig.\ \ref{fig:reqsnr}. Comparison with the expected binary-inspiral detection rates for second-generation ground-based detectors \cite{2010CQGra..27q3001A} suggests that precise tests of GR would have to wait for the much higher rates afforded by third-generation detectors \cite{2012CQGra..29l4013S}. Even pooling together the evidence from all observed events \cite{2010PhRvD..81h4029M} may not help much, reducing the number of required detections by a factor of only a few, because the evidence is dominated by the few loudest sources (see again Sec.\ \ref{sec:decision}). By contrast, space-based observatories such as the LISA concept \cite{lisasciencecase} (or its latest incarnation, the European-led eLISA \cite{2012CQGra..29l4016A}) are not volume-limited for some source classes, and would see some events with large SNRs.

The rest of this paper is organized as follows: in Sec.\ \ref{sec:oddsratio}, I derive the odds ratio in the two cases where the underlying signal is AG and GR; in Sec.\ \ref{sec:decision}, I study the statistics of the AG decision scheme; in Sec.\ \ref{sec:discussion}, I discuss the significance and applications of these results.

\section{AG--GR odds ratio in the high-SNR limit}
\label{sec:oddsratio}

In the following, we let $\theta^i$ be the $m$-dimensional vector of GR parameters, and $\theta^\mu \equiv (\theta^i, \theta^a)$ the vector of AG parameters, which augments $\theta^i$ with the single AG parameter $\theta^a$; the derivation can be extended easily to more AG parameters. We write the true signal as $h_\ag(\theta^\mu_\mathrm{true}) = h_0 + \Delta h$, with $h_0$ a GR signal, and $\Delta h$ the AG correction, with magnitude proportional to $\theta^a_\mathrm{true}$.

In a sufficiently small neighborhood of $\theta^\mu_\mathrm{true}$, the signal can be expanded as $h_\ag(\theta^\mu) = h_0 + \Delta h + \Delta \theta^\mu h_\mu$, with $\Delta \theta^\mu = \theta^\mu - \theta^\mu_\mathrm{true}$ and $h_\mu \equiv \partial h/\partial \theta^\mu$, evaluated at $h_0$. If the SNR is sufficiently large, this approximation is valid throughout the region of parameter space that supports most of the likelihood \cite{2008PhRvD..77d2001V}.

We can now compute the value $P(\ag|s_\ag)$ of the evidence for the AG hypothesis when the data contain an AG signal, $s_\ag = h_\ag(\theta^\mu_\mathrm{true}) + n$. The likelihood can be written as
\begin{equation}
p(s_\ag|\Delta \theta^\mu) = \mathcal{N} e^{-|s_\ag - h(\theta^\mu)|^2/2}
= \mathcal{N} e^{-|n - \Delta \theta^\mu h_\mu|^2/2},
\end{equation}
and it is maximized by $\Delta \theta^\mu_\mathrm{ML} = (G^{-1})^{\mu\nu} (n,h_\nu)$, with $G_{\mu \nu} = (h_\mu,h_\nu)$ the $(m+1)$-dimensional AG \emph{Fisher matrix}. Switching to parameters $\delta \theta^\mu = \Delta \theta^\mu - \Delta \theta^\mu_\mathrm{ML}$ that describe displacements around the maximum, we resum the exponential as
\begin{widetext}
\begin{equation}
p(s_\ag|\delta \theta^\mu) = \mathcal{N} e^{-|n|^2/2
+ (G^{-1})^{\mu\nu} (n,h_\mu)(n,h_\nu)/2 - G_{\mu\nu} \delta \theta^\mu \delta \theta^\nu/2}.
\end{equation}
The evidence follows by integrating out the $\delta \theta^\mu$, which we do under the assumptions of flat priors $p(\theta^\mu) = 1/\Delta \theta^\mu_\mathrm{prior}$ in the relevant region of parameter space, large enough to encompass the Fisher-matrix exponential:
\begin{equation}
P(\ag|s_\ag) = P(\ag) \int p(\theta^\mu) p(s_\ag|\delta \theta^\mu)
= P(\ag) \frac{(2\pi)^{(m+1)/2} \sqrt{|G^{-1}|}}{\prod_\mu \Delta \theta^\mu_\mathrm{prior}}
\, \mathcal{N} e^{-|n|^2/2 + (G^{-1})^{\mu\nu} (n,h_\mu)(n,h_\nu)/2}.
\end{equation}
This expression can be understood as the product of the maximum likelihood (the normalized exponential) with the prior $P(\ag)$ and the Bayesian Occam factor (the fraction), which weighs (by volume) the region of uncertainty for the AG parameters after the observation with the region allowed by their priors. In the high-SNR limit, the posterior region of uncertainty is just the Fisher 1-$\sigma$ ellipsoid, which has volume proportional to $\sqrt{|G^{-1}|}$. The second term in the exponential is the enhancement of likelihood due to overfitting noise: this is a random variable (a function of the noise realization) with expectation value\footnote{\label{foot:prod}From the definition of inner product as $(a,b) = 4 \, \mathrm{Re} \int a^*(f) b(f)/S_n(f) \, df$ and the definition of noise spectral density $S_n$ from $\langle n^*(f) n(f') \rangle = S_n(f) \delta(f - f')/2$, it follows in general that $\langle (n,a)(n,b) \rangle_n = (a,b)$. Then $\langle (G^{-1})^{\mu\nu} (n,h_\mu)(n,h_\nu)\rangle = (G^{-1})^{\mu\nu} G_{\nu\mu} = I^\mu_\mu = m+1$.} equal to $m+1$.

We repeat this computation for the GR hypothesis, expanding the signal as $h_\gr(\theta^i) = h_0 + \Delta \theta^i h_i$, with $\Delta \theta^i = \theta^i - \theta^i_\mathrm{true}$, and integrating over $\delta \theta^i = \Delta \theta^i - (F^{-1})^{ij} (n + \Delta h,h_j)$, with $F_{ij}$ the $m$-dimensional GR Fisher matrix. From the point of view of GR waveforms, $\Delta h$ behaves as an additional noise component. Thus
\begin{equation}
P(\gr|s_\ag) = P(\ag) \frac{(2\pi)^{m/2} \sqrt{|F^{-1}|}}{\prod_i \Delta \theta^i_\mathrm{prior}} \, \mathcal{N} 
e^{-|n + \Delta h|^2/2 + (F^{-1})^{ij} (n + \Delta h,h_i)(n + \Delta h,h_j)/2},
\end{equation}
where $F_{ij} \equiv (h_i,h_j)$ is the $m$-dimensional Fisher matrix.

We can now form the odds ratio $\odds_\ag = P(\ag|s_\ag)/P(\gr|s_\ag)$, using the shorthand $X_\mu \equiv (X,h_\mu)$:
\begin{equation}
\odds_\ag =
\frac{p(\ag)}{p(\gr)} \frac{(2\pi)^{1/2} \sqrt{|G^{-1}|/|F^{-1}|}}{\Delta \theta^\ag_\mathrm{prior}}
e^{[|\Delta h|^2 - (F^{-1})^{ij} \Delta h_i \Delta h_j]/2
+ [(\Delta h,n) - (F^{-1})^{ij} \Delta h_i n_j]
+ [(G^{-1})^{\mu \nu} n_\mu n_\nu - (F^{-1})^{ij} n_i n_j]/2
};
\end{equation}
this expression can be simplified considerably by noting that $(F^{-1})^{ij} h_i (h_j,\cdot)$ acts as the linear projector $P_\gr$ onto the local tangent space of signal derivatives taken with respect to GR parameters, so
\begin{equation}
\begin{gathered}
|\Delta h|^2 - (F^{-1})^{ij} \Delta h_i \Delta h_j = |(1 - P_\gr)\Delta h|^2, \\
(\Delta h,n) - (F^{-1})^{ij} \Delta h_i n_j = ((1 - P_\gr)\Delta h,n);
\end{gathered}
\end{equation}
thus it is only the component $\Delta h_\perp \equiv (1 - P_\gr)\Delta h$ of the AG correction that enters the odds ratio; this is indeed the \emph{residual} that cannot be reabsorbed by shifting the estimated values of the GR parameters, and the larger the $\Delta h_\perp$, the more evidence there is for the AG hypothesis.

The Occam factor and noise-overfitting contributions to the maximum likelihood also bear some simplification: using the block-matrix decomposition of $G_{\mu \nu}$ and its inverse,
\begin{equation}
G_{\mu\nu} = \left(\begin{array}{cc}
F_{ij} & b_i \\
b_j & c
\end{array}\right), \quad
(G^{-1})^{\mu\nu} = \left(\begin{array}{cc}
(F^{-1})^{ij} + (F^{-1})^{ik} b_k b_l (F^{-1})^{lj}/k & -(F^{-1})^{ik} b_k/k \\
-b_k (F^{-1})^{kj}/k & 1/k
\end{array}\right),
\end{equation}
\end{widetext}
where $b_i = (h_i,h_a)$, $c = (h_a,h_a)$, and $k = c - b_i b_j (F^{-1})^{ij}$, we can show that
\begin{equation}
\begin{gathered}
|G_{ij}| = |c F_{ij} - b_i b_j| = |F_{ij}| k, \\
(G^{-1})^{\mu \nu} n_\mu n_\nu - (F^{-1})^{ij} n_i n_j = (\Delta h_\perp,n)^2/|\Delta h_\perp|^2,
\end{gathered}
\end{equation}
so
\begin{equation}
\label{eq:agevidence}
\Oc_\ag =
\frac{p(\ag)}{p(\gr)} \frac{(2\pi)^{1/2} \Delta \theta^a_\mathrm{est}}{\Delta \theta^a_\mathrm{prior}}
\, e^{|\Delta h_\perp|^2/2 + x|\Delta h_\perp| + x^2/2
},
\end{equation}
where $x = (\Delta h_\perp,n)/|\Delta h_\perp|$ is a normal random variable with zero mean and unit variance (see again footnote \ref{foot:prod}),
and $\Delta \theta^a_\mathrm{est} = k^{-1/2}$ is the estimation error for the AG parameter, as given by the corresponding diagonal element of the inverse Fisher matrix $G^{-1}$. Remarkably (if logically), the odds ratio turns out to be a function of the posterior uncertainty and prior range for the additional AG parameter alone.

We can link $\Delta h_\perp$ to the fitting factor FF by finding the $\Delta \theta^i$ that maximizes the normalized match
\begin{equation}
\label{eq:defff}
\ff = \max_{\Delta \theta^i}
\frac{(h_0 + \Delta h,h_0 + \Delta \theta^i h_i)}{|h_0 + \Delta h| \cdot |h_0 + \Delta \theta^i h_i|},
\end{equation}
which is given (unsurprisingly) by $\Delta \theta^i = (F^{-1})^{ij} (\Delta h,h_j)$, and replacing it in Eq.\ \eqref{eq:defff}, yielding
\begin{equation}
\label{eq:fffromperp}
1 - \ff = \frac{1}{2} \frac{|\Delta h_\perp|^2}{|h_0|^2}
= \frac{1}{2} \frac{|\Delta h_\perp|^2}{\snr^2},
\end{equation}
which is valid to $O(\snr^{-4})$. Thus, for fixed FF the odds ratio scales as $\snr^2$, just as it does in the Bayesian decision scheme for the (non)detection of a known signal in noise; for fixed $\snr$ the odds ratio scales as $1 - \ff$, so the odds ratio is larger with stronger and less reabsorbable AG deviations. The effects of detector noise add some statistical fluctuations through the random variable $x$.

This derivation can be repeated with small changes to yield the odds ratio when the data contain a GR signal, $s_\gr = h_\gr(\theta^i_\mathrm{true}) + n$, with $h_\gr(\theta^i_\mathrm{true}) = h_0$, leading to
\begin{equation}
\label{eq:grevidence}
\Oc_\gr = \frac{p(\ag)}{p(\gr)} \frac{(2\pi)^{1/2} \Delta \theta^a_\mathrm{est}}{\Delta \theta^a_\mathrm{prior}}
\, e^{x^2/2},
\end{equation}
where again $x$ is a normal random variable with zero mean and unity variance. Equations \eqref{eq:agevidence}, \eqref{eq:fffromperp}, and \eqref{eq:grevidence} comprise the main novel result of this paper, and in the next section we use them to characterize the statistics of our decision scheme.

\vspace{12pt}
\section{AG--GR decision scheme}
\label{sec:decision}

The distribution of $\odds_\gr$, as implied by the distribution of $x$ through Eq.\ \eqref{eq:grevidence}, determines the \emph{background} of false AG detections for a chosen threshold $\Oc_\thr$, quantified by the false-alarm probability $F = P(\Oc_\gr > \Oc_\thr)$. We choose $\Oc_\thr$ to yield the desired $F$, and evaluate the corresponding efficiency $E = P(\Oc_\ag > \Oc_\thr)$ from Eq.\ \eqref{eq:agevidence}.
Surprisingly, because the ratios of priors $P(\ag)/P(\gr)$ and the Occam factors are the same in $\Oc_\gr$ and $\Oc_\ag$, their only effect is to rescale $\Oc_\thr$, and they cancel out when we compute $E$ as a function of $F$. We can then work with the renormalized odds ratios
\begin{equation}
\begin{aligned}
\Oc'_\gr &= e^{x^2/2}, \\
\Oc'_\ag &= e^{x^2/2 + \sqrt{2} \, x \, \snr_\ag + \snr_\ag^2},
\end{aligned}
\end{equation}
where $\snr_\ag \equiv \snr \sqrt{1 - \ff}$ plays the role of an \emph{effective} SNR for AG detection.

This is not to say that the priors $P(\ag)$ and $P(\gr)$ are unimportant. Indeed, our prior degree of belief in AG sets our requirements for $F$ \cite{2008CQGra..25r4010V}. From basic Bayesian reasoning, the probability that AG is true when it is ``detected'' the odds-ratio decision scheme is
\begin{equation}
\begin{aligned}
P(\ag|\mathrm{detected}) &= \frac{E \times P(\ag)}{
E \times P(\ag)
+ F \times P(\gr)} \\
&= \left(1 + \frac{F}{E} \frac{P(\gr)}{P(\ag)}\right)^{-1};
\end{aligned}
\end{equation}
since GR is so well tested, it seems reasonable that $P(\ag) \ll P(\gr)$; then $F$ must be $\ll P(\ag)$ if we are to believe that we have truly detected AG, because a false alarm is \emph{a priori} much more probable than a true detection.

Combining Eq.\ \eqref{eq:grevidence} with the definition of $F$ and the sampling distribution $p(x) = e^{-x^2/2} / \sqrt{2\pi}$, we obtain
\begin{equation}
\label{eq:f}
F = \mathrm{erfc}\Bigl(\sqrt{\log \Oc'_\thr}\Bigr),
\end{equation}
with $\mathrm{erfc}(z) = 1 - \mathrm{erf}(z)$ the \emph{complementary error function}, defined from the error function\footnote{With this definition, the c.d.f.\ of a normal variable $x$ with zero mean and unit norm is $\mathrm{cdf}(x) = 1/2 (1 + \mathrm{erf}(x/\sqrt{2})).$} $\mathrm{erf}(z) = (2/\sqrt{\pi}) \int_0^z e^{-t^2} dt$. Likewise, combining Eq.\ \eqref{eq:agevidence} with the definition of $E$ and $p(x)$, we find
\begin{widetext}
\begin{equation}
\label{eq:e}
E = \frac{1}{2}\biggl(
\mathrm{erf}\Bigl(-\snr_\ag + \sqrt{\log \Oc'_\thr}\Bigr) - \mathrm{erf}\Bigl(-\snr_\ag - \sqrt{\log \Oc'_\thr}\Bigr)
\biggr).
\end{equation}
Next, we solve Eq.\ \eqref{eq:f} for $\Oc'_\thr$ and replace it in Eq.\ \eqref{eq:e}:
\begin{equation}
\label{eq:roc}
E = 1 - \frac{1}{2}\biggl( \mathrm{erf}\Bigl(-\snr_\ag + \mathrm{erfc}^{-1}(F)\Bigr) - 
\mathrm{erf}\Bigl(-\snr_\ag - \mathrm{erfc}^{-1}(F)\Bigr) \biggr),
\end{equation}
\end{widetext}
where $z = \mathrm{erfc}^{-1}(P)$ is the solution of $\mathrm{erfc}(z) = P$. Solving $E(\snr_\ag) = 1/2$ yields the $\snr_\ag$ required for confident AG detection as a function of $F$, ranging from 2.75 to 4.05 for $F = 10^{-4}$ down to $10^{-8}$. The GW-detection $\snr$ required for AG detection is just $\snr_\ag(F)/\sqrt{1 - \ff}$, and it is plotted in Fig.\ \ref{fig:reqsnr} for $F = 10^{-4}$ and $10^{-8}$. We already discussed the meaning of these curves in Sec.\ \ref{sec:intro}.

An interesting question to ask is what detection rates would be needed in a volume-limited search so that the \emph{loudest} observed signal could be used to detect AG corrections of given $\ff$. In such a search, neglecting cosmological effects for simplicity, source distances are distributed as $p(D) = 3/D_\mathrm{hor} (D/D_\mathrm{hor})^2$, out to the horizon distance $D_\mathrm{hor}$ where sources are detected at the threshold $\snr_\thr$. For $N$ GW detections, the minimum distance is distributed\footnote{Why? Consider first the minimum $x_\mathrm{min}$ among $N$ variables independently and uniformly distributed in $[0,1]$. Its distribution is $p(x_\mathrm{min}) = N (1-x_\mathrm{min})^{N-1}$, since we could pick any of the $N$ as the minimum, and then its probability of being in $[x_\mathrm{min},x_\mathrm{min}+dx]$ is just $dx$ times the probability that the other $N-1$ are in $[x_\mathrm{min},1]$. The minimum $y_\mathrm{min}$ among $N$ variables with distribution $p(y_\mathrm{min})$ follows from the transformation $x = \mathrm{cdf}(y)$, from which $p(y_\mathrm{min}) = p(x_\mathrm{min}) \frac{dx}{dy}|_{y_\mathrm{min}}$.} as $p(D_\mathrm{min}) = 3 N/D_\mathrm{hor} (D/D_\mathrm{hor})^2 (1-(D/D_\mathrm{hor})^3)^{N-1}$, which has median $D_\mathrm{hor} (1 - 2^{-1/N})^{1/3}$. If follows that the median maximum SNR is $\snr_\mathrm{thr} (1 - 2^{-1/N})^{-1/3}$. Setting this equal to $\snr_\ag(F)/\sqrt{1 - \ff}$ and solving for $N$, we obtain the required number of detections, which scales as $(1 - \ff)^{-3/2}$,
and is shown in Fig.\ \ref{fig:reqsnr} on the right-side vertical axis for $\snr_\mathrm{thr} = 8$.

Figuring out what happens if we pool together the evidence from a number of observed events \cite{2010PhRvD..81h4029M,2012PhRvD..85h2003L} of the same kind is a little harder computationally. The odds ratios take forms similar to the one-signal case:
\begin{equation}
\begin{aligned}
\Oc'_\gr &= e^{\sum_i x_i^2/2}, \\
\Oc'_\ag &= e^{\sum_i x_i^2/2 + \sqrt{2} \sum_i x_i \, \snr_{\ag,i} + \snr_{\ag,i}^2},
\end{aligned}
\end{equation}
where the $x_i$ are independently distributed normal variables with zero mean and unit variance, and the $\snr_{\ag,i}$ are the effective AG-detection SNRs for the individual observations. Here I limit myself to a small Monte Carlo exploration: assuming for simplicity that the FF is the same for all the sources, and taking the median over all $\{\snr_{\ag,i}\}$ realizations in a volume-limited search with $\snr_\mathrm{thr} = 8$, I find that with $F = 10^{-4}$ we need $\sim 9$/$200$/$4,500$ observations to detect AG with $1 - \ff = 10^{-2}$/$10^{-3}$/$10^{-4}$, to be compared with $\sim 28$/$900$/$30,000$ using evidence from the loudest source alone. Essentially, because SNRs are distributed as $1/\snr^4$, the Bayesian-inference problem is dominated by a few very loud events, and there are not very many of those for moderate detection rates. (However, this conclusion differs from the findings of Ref.\ \cite{2012PhRvD..85h2003L}, and it would be interesting to understand why.)

\section{Discussion}
\label{sec:discussion}


In this paper I have shown that, under the assumptions of strong signals and Gaussian detector noise, the prospects for detecting alternative-gravity corrections to General Relativity can be characterized very simply. A single number, the fitting factor FF between the GR and AG waveform families, determines the source SNR required for the alternative-gravity hypothesis to be favored in a decision scheme based on the Bayesian odds ratio (see Fig.\ \ref{fig:reqsnr}).

This happens because the FF is an SNR-independent measure of the strength of the AG corrections $\Delta h_\perp$ that cannot be reabsorbed by changing the GR source parameters from their true values. The GR parameters are not known \emph{a priori}, but must be determined from the same observation, so such ``reabsorbable'' AG effects cannot be detected positively, and they would result in a ``fundamental bias'' \cite{2009PhRvD..80l2003Y} on the GR parameters if AG is true, but post-detection parameter estimation is performed with GR model templates. 
In Ref.\ \cite{2011PhRvD..84f2003C}, Cornish and colleagues call such errors ``stealth bias'' if they are comparable to or larger than the noise-induced statistical errors in the GR parameters, and yet AG cannot be detected positively. In the terms of this paper, stealth bias corresponds to FF very close to one and AG-induced errors $(F^{-1})^{ij} (\Delta h,h_j)$ that are large compared to the Fisher-matrix statistical errors $\sqrt{(F^{-1})^{ii}}$.

My formalism can also be applied to other contexts\footnote{I thank Ilya Mandel for pointing this out and providing these examples.} where we need to decide between a simpler model and one with additional parameters, such as binary inspirals of nonspinning vs.\ spinning compact objects, orbit-aligned vs.\ precessing spins, or point-like vs.\ extended-object dynamics.

My formulas cannot predict what happens when the high-SNR, linearized-parameter approximation is not warranted; whether that is the case can be determined using the test described in Sec.\ VI of Ref.\ \cite{2008PhRvD..77d2001V}. At low SNRs, full-fledged Monte Carlo integration \cite{2011PhRvD..84f2003C,2011PhRvD..83h2002D,2012PhRvD..85h2003L} would be required for accurate predictions, although the FF formula could be used as a preliminary step, and its comparison with the full result would be very instructive. I note however that it is for the strongest signals that GR-by-GW tests become most interesting, and that the results discussed above would persist as the leading-order contributions to the evidence (see again \cite{2008PhRvD..77d2001V}, Sec.\ VII).

Beyond the statistical characterization of the tests, we should always ask ourselves \emph{what it is} that we could really detect, and whether \emph{we should really believe} a positive AG detection if we get it. These are very hard questions, and here I offer only some qualitative considerations that should be kept in mind whenever we discuss the sensitivity of GR-by-GW tests.

First, it seems evident that a test based on matching AG corrections of a certain functional form $\Delta h$ would only be sensitive to non-GR effects that have nonzero projection along $\Delta h$. (For instance, AG waveforms with additional phasing parameters would not be sensitive to amplitude corrections.) Now, both the consistency checks based on altering PN coefficients and the ppE framework consider rather general corrections, so it may be hard to imagine that the waveform imprint from any reasonable AG theory would be fully orthogonal to them. 
Indeed, Ref.\ \cite{2012PhRvD..86b2004C} argues that for quasicircular binary inspirals, the well-posedness of the initial-value problem restricts possible phasing terms to frequency powers $f^{n/3}$ (where $n$ can be negative), which could be covered in the ppE scheme.
However, if the projection is small, the resulting $1 - \ff$ would be strongly reduced, and the test would be sensitive only to much larger effects.

Second, any positive detection of an AG correction $\Delta h$ could also be explained as one of many \emph{systematic} waveform corrections \cite{2011PhRvD..84b4032K} that have nonzero projection along $\Delta h$, such as the effects of detector calibration and non-Gaussian detector noise, of standard-GR physics not included in the waveforms (spins, eccentricity, higher-PN terms), and of astrophysical perturbations (accretion disks, three-body systems). All of these effects should be considered \emph{a priori} more likely than a modification of the extensively well tested GR, so they must be controlled by including them explicitly in the GR model, or at least by establishing that they are sufficiently orthogonal to AG corrections. On the plus side, instrumental systematics would be different for the same signal as observed in multiple detectors, and GR-theoretical and astrophysical systematics would be different for multiple signals from similar sources, which would help discriminate AG corrections \cite{2011PhRvL.107q1103Y}. Nevertheless, preliminary claims of sensitivity to specific AG corrections may be overoptimistic, because $\Delta h$ could be largely reabsorbed by systematic effects that are initially neglected.

Testing GR with GWs remains one of the exciting frontiers of GW astronomy, but appropriate caution is needed to provide the proper context for current and future investigations, and to  allocate research effort wisely as we move toward the GW detection era. Computing some FFs will help.

\acknowledgments I am grateful to Walter Del Pozzo, Marc Favata, Ilya Mandel, Chris Messenger, Reinhard Prix, and (especially) Curt Cutler, Nico Yunes, and the anonymous PRD referee for useful discussions and suggestions. 
This work was carried out at the Jet Propulsion Laboratory, California Institute of Technology, under contract with the National Aeronautics and Space Administration. MV was supported by the LISA Mission Science Office and by the JPL RTD program. Government sponsorship acknowledged. Copyright 2012 California Institute of Technology.

\bibliography{paper}

\end{document}